\documentclass[10pt]{article}
\usepackage{amsfonts}
\usepackage{epsfig}
\usepackage{xspace}

\newlength{\StrutWidth}
\newcommand{\VarUpStrut}[1]{\rule{\StrutWidth}{#1}}
\newcommand{\VarDownStrut}[1]{\rule[-#1]{\StrutWidth}{#1}}
\newcommand{\NoMathAfterHLS}{\VarUpStrut{2.4ex}}
\newcommand{\MathAfterHLS}{\VarUpStrut{3.0ex}}
\newcommand{\MathBetweenRowsS}{\VarUpStrut{2.4ex}}
\newcommand{\BeforeHLS}{\VarDownStrut{1.2ex}}

\newtheorem{Theorem}{Theorem}
\newcommand{\ThAttrTo}[1]{\emph{#1}\\}
\newtheorem{Lemma}{Lemma}
\newtheorem{Definition}{Definition}
\newtheorem{Proposition}{Proposition}

\newcommand{\DefEmph}[1]{\emph{#1}}
\newcommand{\Proof}{{\sc Proof:~}}
\newcommand{\BProof}{{\sc Proof:~}}
\newcommand{\EProof}{\ensuremath{\diamond}}
\newcommand{\Warning}{{\sc Warning: }}

\newcommand{\BTh}{\begin{Theorem}}
\newcommand{\ETh}{\end{Theorem}}
\newcommand{\BLemma}{\begin{Lemma}}
\newcommand{\ELemma}{\end{Lemma}}
\newcommand{\BDef}{\begin{Definition}}
\newcommand{\EDef}{\end{Definition}}
\newcommand{\BProp}{\begin{Proposition}}
\newcommand{\EProp}{\end{Proposition}}
\newcommand{\BFact}{\begin{Fact}}
\newcommand{\EFact}{\end{Fact}}

\newcommand{\BEx}{\begin{quotation}\begin{small}\textsc{Example:~}}
\newcommand{\BExCont}{\begin{quotation}\begin{small}\textsc{Example} (continued):~}
\newcommand{\EEx}{\end{small}\end{quotation}}

\newcommand{\BCenter}{\begin{center}}
\newcommand{\BCentre}{\begin{center}}
\newcommand{\ECenter}{\end{center}}
\newcommand{\ECentre}{\end{center}}
\newcommand{\ili}[1]{\mbox{\it (#1) }\,}
\newcommand{\BItem}{\begin{itemize}}
\newcommand{\EItem}{\end{itemize}}

\newcommand{\BTItem}{\begin{description}}
\newcommand{\ETItem}{\end{description}}

\newcommand{\BEqn}{\begin{equation}}
\newcommand{\EEqn}{\end{equation}}

\newcommand{\BEqnA}{\begin{eqnarray}}
\newcommand{\EEqnA}{\end{eqnarray}}

\newcommand{\Prob}{\ensuremath{\mathrm{Prob}}}
\newcommand{\Trace}{\ensuremath{\mathrm{Tr}}}

\newcommand{\POVM}{\mbox{\sc povm}\xspace}
\newcommand{\PVM}{\mbox{\sc pvm}\xspace}

\newcommand{\BMS}{\mbox{\sc bms}\xspace}

\renewcommand{\epsilon}{\varepsilon}
\renewcommand{\phi}{\varphi}
\newcommand{\Hash}{\mbox{\#}}
\newcommand{\Zero}{\mbox{$0$}\xspace}
\newcommand{\One}{\mbox{$1$}\xspace}

\newcommand{\Star}{\ast}
\newcommand{\Sqrt}[1]{\sqrt{#1}}

\newcommand{\LAcc}{\ensuremath{\{}}
\newcommand{\RAcc}{\ensuremath{\}}}
\newcommand{\LA}{\ensuremath{\langle}}
\newcommand{\RA}{\ensuremath{\rangle}}

\newcommand{\BinC}[2]{{#1 \choose #2}}

\newcommand{\HS}{\ensuremath{\mathcal{H}}}
\newcommand{\Tensor}{\otimes}
\newcommand{\Vec}[2]{{#1 \choose #2}}
\newcommand{\Matrix}[2]{
    \left (
    \begin{array}{cc}
    #1 \VarUpStrut{2.4ex}\\
    #2 \VarUpStrut{2.4ex}
    \end{array}
    \right ) }
\newcommand{\Ket}[1]{\ensuremath{ | #1 \RA}}

\newcommand{\Proj}[1]{\ensuremath{ | #1 \RA \LA #1 | }}

\newcommand{\Inner}[2]{\ensuremath{ \LA #1 | #2 \RA}}
\newcommand{\LSSR}[2]{\ensuremath{\LA #1 | #2 | #1 \RA }}

\newcommand{\Abs}[1]{\ensuremath{| #1 |}}

\newcommand{\Id}{\ensuremath{{\rm Id}}}

\newcommand{\CalE}{\ensuremath{\mathcal{E}}\xspace}
\newcommand{\CalH}{\ensuremath{\mathcal{H}}}\xspace
\newcommand{\SVT}{\ensuremath{T}\xspace}
\newcommand{\SVX}{\ensuremath{X}\xspace}

\newcommand{\SetX}{\ensuremath{\mathcal{X}}\xspace}

\newcommand{\ShE}{\ensuremath{\mathtt{H}}\xspace}
\newcommand{\BShE}{\ensuremath{\ShE_2}}\xspace
\newcommand{\ShI}{\ensuremath{\mathtt{I}}\xspace}
\newcommand{\A}{\ensuremath{\sf A}\xspace}
\newcommand{\B}{\ensuremath{\sf B}\xspace}

\newcommand{\ZO}{\ensuremath{\{0,1\}}}
 
\newcommand{\Eq}{\mbox{$=$}}

\newcommand{\IsDef}{\ensuremath{\stackrel{\mathrm{def}}{=}}}

\newcommand{\Is}{\Eq}
\newcommand{\Orth}{\ensuremath{\bot}}

\newcommand{\And}{\ensuremath{\wedge}}

\newcommand{\Half}{\mbox{$\frac{1}{2}$}\xspace}
\newcommand{\Quart}{\mbox{$\frac{1}{4}$}\xspace}

\newcommand{\TwoThird}{\mbox{$\frac{2}{3}$}\xspace}
\newcommand{\OneSixth}{\mbox{$\frac{1}{6}$}\xspace}

\newcommand{\WebPage}[1]{\textsf{\footnotesize #1}}
\newcommand{\MyRef}[1]{\ref{#1}}
\newcommand{\MyLabel}[1]{\label{#1}}
\newcommand{\RefEq}[1]{(\MyRef{Eq:#1})}
\newcommand{\LabelEq}[1]{\MyLabel{Eq:#1}}
\newcommand{\MyCit}[1]{{\cite{#1}}}
\newcommand{\MyCitt}[2]{\cite[{#2}]{#1}}

\newcommand{\CPE}{\ensuremath{\mathtt{PE}}\xspace}
\newcommand{\CK}{\ensuremath{\mathtt{K}}\xspace}
\newcommand{\CB}{\ensuremath{\mathtt{B}}\xspace}
\newcommand{\CSD}{\ensuremath{\mathtt{SD}}\xspace}

\newcommand{\QPE}{\ensuremath{\texttt{\textit{PE}}}\xspace}
\newcommand{\QK}{\ensuremath{\texttt{\textit{K}}}\xspace}
\newcommand{\QB}{\ensuremath{\texttt{\textit{B}}}\xspace}
\newcommand{\QSD}{\ensuremath{\texttt{\textit{SD}}}\xspace}
\newcommand{\Overlap}{\ensuremath{\textit{overlap}}\xspace}

\newcommand{\CardX}{\ensuremath{m}\xspace}
\newcommand{\PZE}{\ensuremath{p_0(\CalE)}\xspace}
\newcommand{\PIE}{\ensuremath{p_1(\CalE)}\xspace}
\newcommand{\KPsi}[1]{\ensuremath{\Ket{\psi_{#1}}}\xspace}
\newcommand{\Rho}[1]{\ensuremath{\rho_{#1}}}
\newcommand{\EOpt}{\ensuremath{\mathcal{E}^{\Star}}\xspace}
\newcommand{\SumX}{\ensuremath{\sum_{x \in \SetX}}}
\newcommand{\Hor}{\ensuremath{\sf H}\xspace}
\renewcommand{\Vert}{\ensuremath{\sf V}\xspace}
\newcommand{\AllMmnts}{\ensuremath{\mathcal{M}}}

\newcommand{\LHS}{left hand side\xspace}
\newcommand{\aaBhCo}{Bhattacharyya coefficient\xspace}
\newcommand{\aaCrh}{cryptograph\xspace}

\newcommand{\aaDistn}{distribution\xspace}
\newcommand{\aaDistns}{distributions\xspace}

\newcommand{\aaDistrns}{distributions\xspace}
\newcommand{\aaDisty}{distinguishability\xspace}
\newcommand{\aaDMs}{density matrices\xspace}
\newcommand{\aaDM}{density matrix\xspace}
\newcommand{\aaEyI}{exponentially indistinguishable\xspace}
\newcommand{\aaEI}{exponential indistinguishability\xspace}

\newcommand{\aaIff}{if and only if\xspace}
\newcommand{\aaIndiste}{indistinguishable\xspace}
\newcommand{\aaIndisty}{indistinguishability\xspace}

\newcommand{\aaKD}{Kolmogorov distance\xspace}

\newcommand{\aaMmnt}{measurement\xspace}

\newcommand{\aaOrth}{orthogonal\xspace}
\newcommand{\aaPD}{probability distribution\xspace}
\newcommand{\aaPDs}{probability distributions\xspace}
\newcommand{\aaPOE}{probability of error\xspace}

\newcommand{\aaPE}{probability of error\xspace}
\newcommand{\aaProbs}{probabilities\xspace}
\newcommand{\aaProby}{probability\xspace}

\newcommand{\aaPry}{probability\xspace}

\newcommand{\aaSD}{Shannon distinguishability\xspace}
\newcommand{\aaSV}{stochastical variable\xspace}

\newcommand{\ie}{i.e.,\xspace}

\title{Cryptographic Distinguishability Measures for Quantum
Mechanical States}
\author{\protect Christopher A. Fuchs$^{(1)}$ and Jeroen van de
Graaf$^{(2)}$
\medskip \\
$^{(1)}$ Bridge Laboratory of Physics, 12-33\\
California Institute of Technology\\
Pasadena, California, 91125, U.~S.~A.\\
$^{(2)}$ D\'epartement IRO, Universit\'e de Montr\'eal\\
C.~P. 6128, Succursale centre-ville,\\ Montr\'eal, Canada H3C 3J7}
\date{3 April 1998}
 
\begin{document}
 
\maketitle

\begin{abstract}

This paper, mostly expository in nature, surveys four measures of
distinguishability for quantum-mechanical states.  This is done
from the point of view of the cryptographer with a particular eye
on applications in quantum cryptography.  Each of the measures
considered is rooted in an analogous classical measure of
distinguishability for probability distributions:  namely, the
probability of an identification error, the Kolmogorov distance, the
Bhattacharyya coefficient, and the Shannon distinguishability (as
defined through mutual information).  These measures have a long
history of use in statistical pattern recognition and classical
cryptography.  We obtain several inequalities that relate the
quantum distinguishability measures to each other, one of which may
be crucial for proving the security of quantum cryptographic key
distribution.  In another vein, these measures and their connecting
inequalities are used to define a {\em single\/} notion of
cryptographic exponential indistinguishability for two families of
quantum states.  This is a tool that may prove useful in the
analysis of various quantum cryptographic protocols.

\end{abstract}

%%%%%%%%%%%%%%%%%%%%%%%%%%%%%%%%%%%%%%%%%%%%%%%%%%%%%%%%%%%%%%%%%%%%%%
\section{Introduction}

The field of quantum cryptography is built around the singular idea
that physical information carriers are always quantum mechanical.
When this idea is taken seriously, new possibilities open up within
cryptography that could not have been dreamt of before.  The most
successful example of this so far has been quantum cryptographic
key distribution.  For this task, quantum mechanics supplies a
method of key distribution for which the security against
eavesdropping can be assured by physical law itself.  This is
significant because the legitimate communicators then need make no
assumptions about the computational power of their opponent.

Common to all quantum cryptographic problems is the way information
is encoded into quantum systems, namely through their quantum-%
mechanical states.  For instance, a $0$ might be encoded into
a system by preparing it in a state $\rho_0$, and a $1$ might
likewise be encoded by preparing it in a state $\rho_1$.  The choice
of the particular states in the encoding will generally determine
not only the ease of information retrieval by the legitimate users,
but also the inaccessibility of that information to a hostile
opponent.  Therefore, if one wants to model and analyze the
cryptographic security of quantum protocols, one of the most basic
questions to be answered is the following.  What does it mean for
two quantum states to be ``close'' to each other or ``far'' apart?
Giving an answer to this question is the subject of this paper.
That is, we shall be concerned with defining and relating various
notions of ``distance'' between two quantum states.

Formally a quantum state is nothing more than a square matrix of
complex numbers that satisfies a certain set of supplementary
properties.  Because of this, any of the notions of distance between
matrices that can be found in the mathematical literature would do
for a quick fix.  However, we adhere to one overriding criterion for
the ``distance'' measures considered here.  The only physical means
available with which to distinguish two quantum states is that
specified by the general notion of a quantum-mechanical measurement.
Since the outcomes of such a measurement are necessarily
indeterministic and statistical, only measures of ``distance'' that
bear some relation to statistical-hypothesis testing will be
considered.  For this reason, we prefer to call the measures
considered herein {\em distinguishability measures\/} rather than
``distances.''

In this paper, we discuss four notions of distinguishability that are
of particular interest to cryptography:  the probability of an
identification error, the Kolmogorov distance (which turns out to
be related to the standard trace-norm distance), the Bhattacharyya
coefficient (which turns out to be related to Uhlmann's ``transition
probability''), and the Shannon distinguishability (which is defined
in terms of the optimal mutual information obtainable about a
state's identity).  Each of these four distinguishability measures
is, as advertised, a generalization of a distinguishability measure
between two probability distributions.

Basing the quantum notions of distinguishability upon classical
measures in this way has the added bonus of easily leading to
various inequalities between the four measures.  In particular, we
establish a simple connection between the \aaPOE and the trace-norm
distance.  Moreover, we derive a very simple upper bound on the \aaSD
as a function of the trace-norm distance:
$\QSD(\rho_0,\rho_1) \le \Half \Trace \Abs{\rho_0-\rho_1}$.
(The usefulness of this particular form for the bound was
realized while one of the authors was working on \MyCit{BihamBoBrGrMo97},
where it is used to prove security of quantum key
distribution for a general class of attacks.)  Similarly, we can bound
the quantum \aaSD by functions of the quantum Bhattacharrya
coefficient.

In another connection, we consider an application of these
inequalities to protocol design.
In the design of cryptographic protocols, one often defines a
\emph{family} of protocols parameterized by a
\emph{security parameter}, $n$---where this number denotes the length
of some string, the number of rounds, the number of photons, etc.
Typically the design of a good protocol requires that the probability
of cheating for each participant vanishes exponentially fast, \ie is
of the order $O(2^{-n})$, as $n$ increases.  As an example, one
technique is to compare the protocol implementation (the family of
protocols) with the \emph{ideal protocol specification} and to prove
that these two become exponentially indistinguishable\footnote{This
notion is more commonly called \emph{statistical indistinguishability}
in the cryptographic literature.  However, since the word
``statistical'' is likely to already be overused in this paper, we
prefer ``exponential.''} \MyCit{GoldwasserMiRa89, Beaver91}.

To move this line of thought into the quantum regime, it is natural
to consider two families of quantum states
parameterized by $n$ and to require that the \aaDisty between the two
families vanishes exponentially
fast.  A priori, this exponential convergence could depend upon which
\aaDisty measure is chosen---after all the quantum-mechanical
measurements optimal for each distinguishability measure can be quite
different.  However, with the newly derived
inequalities in hand, it is an easy matter to show that exponential
indistinguishability with respect to one measure implies exponential
indistinguishability with respect to each of the other four measures.
In other words, these four notions are equivalent, and it is
legitimate to speak of a single, unified {\em exponential
indistinguishability\/} for two families of quantum states.

The contribution of this paper is three-fold.  In the first place,
even though some of the quantum inequalities derived here are minor
extensions of classical inequalities that have been known for some
time, many of the classical inequalities are scattered
throughout the literature in fields of research fairly remote from
the present one.  Furthermore, though elements of this work can also
be found in \MyCit{FuchsPhD}, there is presently no paper
that gives a systematic overview of quantum distinguishability
measures from the cryptographer's point of view.
In the second place, some of the inequalities in Section 6
are new, even within the classical regime. In the third place, a
canonical definition for quantum exponential
indistinguishability is obtained.  The applications of this notion
may be manifold within quantum cryptography.

The structure of the paper is as follows.  In the following section
we review a small bit of standard probability theory, mainly to
introduce the setting and notation.  Section 3 discusses density
matrices and measurements, showing how the combination of the two
notions leads to a probability distribution.  In Section 4 we discuss
four measures of distinguishability, first for classical probability
distrubitions, then for quantum-mechanical states.
After a short summary in Section 5, we discuss several inequalities,
again both classically and quantum mechanically.  In Section 6 these
inequalities are applied to proving a theorem about exponential
indistinguishability.  Section 7 discusses an application of this
notion---in particular, we give a simple proof of a theorem in
\MyCit{BennettMoSm96} that the Shannon distinguishability of the
parity (\ie the overall exclusive-or) of a quantum-bit-string
decreases exponentially with the length of the string.  Moreover,
the range of applicability of the theorem is strengthened in the
process.

This paper is aimed primarily at an audience of computer scientists,
at cryptographers in particular, with some small background
knowledge of quantum mechanics.  Readers needing a more systematic
introduction to the requisite quantum theory should consult
Hughes~\MyCit{HughesBook} or Isham~\MyCit{IshamBook}, for instance.
A very brief introduction can be found in the appendix of
\MyCit{BrassardCrJoLa93}.

%%%%%%%%%%%%%%%%%%%%%%%%%%%%%%%%%%%%%%%%%%%%%%%%%%%%%%%%%%%%%%%%%%%%%%
\section{Probability distributions} \MyLabel{Sec:PDs}

Let $\SVX_0$ be a stochastic variable over a finite set $\SetX$.
Then we can define $ p_0(x) \IsDef \Prob [\SVX_0 = x]$,
so $\SVX_0$ induces a probability distribution $p_0$ over $\SetX$.
Let $p_1$ be defined likewise.
Of course, $\sum_{x \in X} p_t(x) = 1$ for $t\Is 0, 1$.
After relabeling the outcomes $x_1, x_2, x_3, \ldots x_\CardX $
to $1, 2, 3, \ldots \CardX$ we get: %\\[0.5ex]

\BCenter
\begin{tabular}{| l | l || l l l l l |}
\hline \NoMathAfterHLS
\BeforeHLS
        & $$   & $x \Eq 1$        & $x \Eq 2$        & $x \Eq 3$ &
                          \ldots  & $x \Eq \CardX $  \\ \hline \hline

\NoMathAfterHLS
$\SVX_0$ & $\pi_0 = \Half$ & $p_0(1)$ & $p_0(2)$ & $p_0(3)$ &
                            \ldots  & $p_0(\CardX)$  \\ %%\hline

\MathBetweenRowsS \BeforeHLS
$\SVX_1$ & $\pi_1 = \Half$ & $p_1(1)$ & $p_1(2)$ & $p_1(3)$ &
                            \ldots  & $p_1(\CardX)$  \\ \hline
\end{tabular}\\[1ex]
\ECenter

Here $\pi_0$ and $\pi_1$ are the \emph{a priori} probabilities
of the two stochastic variables; they sum up to $1$.
Throughout this paper we take $\pi_0 = \pi_1 = \Half$.
(Even though much of our analysis could be extended to the case
$\pi_0 \ne \pi_1 \ne \Half$, it seems not too relevant for the
questions addressed here.)
Two \aaDistrns are \emph{equivalent} (\ie \emph{\aaIndiste{}})
if \( p_0(x) = p_1(x) \) for all $x \in X$, and they are
\emph{\aaOrth{}} (\ie \emph{maximally indistinguishable})
if there exists no $x$ for which both $p_0(x)$ and $p_1(x)$ are
nonzero.

Observe that $p_t(x)$ denotes the conditional probability
that $\SVX \Is x$ \emph{given} that $\SVT \Is t$,
written as $\Prob[\SVX \Is x | \SVT \Is t]$.
So the joint \aaProby is half that value:
\BEqnA
\Prob[\SVX \Is x \And \SVT \Is t] & = &
 \Prob [ \SVT \Is t ] \ \Prob[ \SVX \Is x | \SVT \Is t] \\
& = & \pi_t p_t(x) \\
& = & \Half p_t(x)\;.
\EEqnA

We define the conditional probability
$r_t(x) := \Prob[\SVT \Is t | \SVX \Is x]$,
and the probability that $\SVX \Is x$
\emph{regardless} of $t$, that is,
$p(x) := \Prob [ \SVX \Is x ]$.
Using Bayes' Theorem we get:
\BEqnA
r_t(x) & = & \Prob [ \SVT \Is t | \SVX \Is x ] \\
& = & \Prob [ \SVT \Is t ] \ \Prob[ \SVX \Is x | \SVT \Is t] \slash
     \Prob [ \SVX \Is x ] \\
& = & \Half p_t(x) \slash p(x)
\EEqnA
Observe that $r_0(x) + r_1(x) = 1$ for all $x$.
Using
$p(x)$ and $r_t(x)$
we can represent the situation also in the following way:
\BCenter
\begin{tabular}{| l | l || l l l l l |}
\hline \NoMathAfterHLS
\BeforeHLS
          & $$   & $x \Eq 1$        & $x \Eq 2$        & $x \Eq 3$ &
                            \ldots  & $x \Eq \CardX $  \\ \hline
\NoMathAfterHLS \BeforeHLS
$\SVX$ & $$ & $p(1)$ & $p(2)$ & $p(3)$ &
                            \ldots  & $p(\CardX)$  \\ \hline \hline
 
\NoMathAfterHLS
$\SVX_0$ & $\pi_0 = \Half$ & $r_0(1)$ & $r_0(2)$ & $r_0(3)$ &
                            \ldots  & $r_0(\CardX)$  \\ %%\hline
 
\MathBetweenRowsS \BeforeHLS
$\SVX_1$ & $\pi_1 = \Half$ & $r_1(1)$ & $r_1(2)$ & $r_1(3)$ &
                            \ldots  & $r_1(\CardX)$  \\ \hline
\end{tabular}\\[1ex]
\ECenter

\section{Density matrices and measurements}

Recall that a quantum state is said to be a \emph{pure} state if
there exists some (fine-grained) measurement that can confirm this
fact with probability 1.  A pure states can be represented by a
normalized vector $|\psi\rangle$ in an $N$-dimensional Hilbert space,
\ie a complex vector space with inner product.  Alternatively it can
be represented by a projection operator $|\psi\rangle\langle\psi|$
onto the rays associated with those vectors.  In this paper $N$ is
always taken to be finite.

Now consider the following preparation of a quantum system: \A flips
a fair coin and, depending upon the outcome, sends one of two
different pure states $\KPsi{0}$ or $\KPsi{1}$ to \B.  Then the
``pureness'' of the quantum state is ``diluted'' by the classical
uncertainty about the resulting coin flip.  In this case, no
deterministic fine-grained measurement generally exists for
identifying \A's exact preparation, and the quantum state is said to
be a \emph{mixed} state.  \B's knowledge of the system---that is, the
source from which he draws his predictions about any potential
measurement outcomes---can now no longer be represented by a vector
in a Hilbert space.  Rather, it must be described by a {\em density
operator\/} or {\em density matrix\/}\footnote{In general, we shall
be fairly lax about the designations ``matrix'' and ``operator,''
interchanging the two rather freely.  This should cause no trouble
as long as one keeps in mind that all operators discussed in this
paper are linear.} formed from a statistical average of the
projectors associated with \A's possible fine-grained preparations.
\BDef
\ThAttrTo{(see for instance \MyCit{SudberyBook, IshamBook, PeresBook})}
A \DefEmph{density matrix} $\rho$ is an $N$$\times$$N$ matrix
with unit trace that is Hermitian (i.e. $\rho = \rho^\dag$) and
positive semi-definite
(i.e., $ \LSSR{\psi}{\rho} \geq 0 $ for all $\psi \in \CalH$).
\EDef

\BEx
Consider the case where \A prepares either a horizontally or a
vertically polarized photon. We can choose a basis such that
\( \Ket{\Hor} = \Vec{1}{0} \)
and
\( \Ket{\Vert} = \Vec{0}{1} \).
Then \A's preparation is perceived by \B as the mixed state
\BEqn
\Half \Proj{\Hor} +  \Half \Proj{\Vert} = 
\Half \Matrix{1 & 0}{0 & 0} +
\Half \Matrix{0 & 0}{0 & 1} =
%%  \Matrix{\Half & 0}{0 & \Half},
\Matrix{1/2 & 0}{0 & 1/2},
\EEqn
which is the ``completely mixed state''.

Note that the same \aaDM will be obtained if \A prepares an equal
mixture of left-polarized and right-polarized photons.  In fact, any
equal mixture of two orthogonal pure states will yield the same
density matrix.
\EEx

Any source of quantum samples (that is, any imaginary \A who
secretly and randomly prepares quantum states according to some
probability distribution) is called an \emph{ensemble}.  This can be
viewed as the quantum counterpart of a stochastic variable.  A
density matrix completely describes \B's knowledge of the sample.
Two different ensembles with the same density matrix are
indistinguishable as far as \B is concerned; when this is the case,
there exists no measurement that can allow \B a decision between
the ensembles with probability of success better than chance.

The fact that a density matrix describes \B's
{\em a priori\/} knowledge implies that additional classical
information can change that density matrix.  This is so, even when no
measurement is performed and the quantum system remains untouched.
Two typical cases of this are: (1)  when \A reveals to \B information
about the the outcome of her coin toss, or (2) when \A and \B share
quantum entanglement (for example Einstein-Podolsky-Rosen, or EPR,
particles), and \A sends the results of some measurements she
performs on her system to \B.  Observe that, consequently, a density
matrix is subjective in the sense that it depends on what \B knows.

\BExCont\ 
\ili1 Suppose that, after \A has sent an equal mixture of
$\Ket{\Hor}$ and $\Ket{\Vert}$, she reveals to \B that for that
particular sample she prepared $\Ket{\Vert}$.
Then \B's \aaDM changes, as far as he is concerned, from 
\BEqn \LabelEq{ExDM}
%%  \Matrix{\Half & 0}{0 & \Half}
\Matrix{1/2 & 0}{0 & 1/2}
\quad \mathrm{to} \quad
\Matrix{0 & 0}{0 & 1}.
\EEqn

\ili2
An identical change happens in the following situation: \A prepares
two EPR-correlated photons in a combined pure state
\BEqn
\Ket{\Psi^-}=\frac{1}{\Sqrt{2}}\,\big(\Ket{\Hor}\Ket{\Vert}-
\Ket{\Vert}\Ket{\Hor}\big)\;,
\EEqn
known as the singlet state.  Following that, she sends one of the
photons to \B.  As far as \B is concerned, his photon's polarization
will be described by the completely mixed state.  On the other hand,
if \A and \B measure both photons with respect to the same
polarization (vertical, eliptical, etc.), we can predict from
the overall state that their measurement outcomes will be
anti-correlated.  So if, upon making a measurement, \A finds that her
particle is horizontally polarized (i.e., $\Ket{\Hor}$) and she tells
this to \B, then \B{}'s \aaDM will change according to \RefEq{ExDM}.
\EEx

As an aside, it is worthwhile to note that physicists sometimes
disagree about whether the density matrix should be regarded as
\emph{the} state of a system or not.  This, to some extent, can
depend upon one's interpretation of quantum mechanics.  Consider, for
instance, the situation where \B has not yet received the additional
classical information to be sent by \A.  What is the state of his
system?  A pragmatist might answer that the state is simply described
by \B's density matrix.   Whereas a realist might argue that 
the state is really something different, namely one of the pure
states that go together to form that density matrix:  \B is merely
ignorant of the ``actual'' state.  For discussion of this topic
we refer the reader to \MyCit{IshamBook,Mermin96}.  Here we leave
this deep question unanswered and adhere to the pragmatic approach,
which, in any case, is more relevant from an information-theoretical
point of view.

%% DM and MMNTS
Now let us describe how to compute the \aaPry of a certain \aaMmnt
result from the \aaDM.  Mathematically speaking, a density matrix
$\rho$ can be regarded as an object to which we can
apply another operator $E_x$ to obtain a probability.  In particular,
taking the trace of the product of the two matrices yields the \aaPry
that the \aaMmnt result is $x$ given that the state was $\rho$,
\ie
$ \Prob [{\rm result} \Is x | {\rm state} \Is \rho] =$
$\Trace (\rho E_x)$.
Here the $x$ serves as a label, connecting the operator $E_x$ and the
outcome $x$, but otherwise has no specific physical meaning.
(This formula may help the reader understand the designation
``density operator'':  it is required in order to obtain a
probability density function for the possible measurement outcomes.)

Most generally, a quantum-mechanical \emph{measurement} is described
formally by a collection (ordered set) of operators, one for each
outcome of the measurement.

\BDef
\ThAttrTo{(see \MyCit{PeresBook})}
Let $\CalE = \LA E_1, \ldots , E_\CardX \RA$
be  a  collection (ordered set) of operators such that
\ili1 all the $E_x$ are positive semi-definite operators, and
\ili2 $\sum_x E_x = \Id$, where $\Id$ is the identity operator.
Such a collection specifies a \DefEmph{Positive Operator-Valued
Measure} (\POVM) and corresponds to the most general type of
measurement than can be performed on a quantum system.

Applying a \POVM to a system whose state is described
by a density matrix $\rho$ results in a \aaPD according to:
\BEqn \LabelEq{TraceStateOp}
\Prob [{\rm result} = x | {\rm state} = \rho] =
\Trace (\rho E_x)
\EEqn
where $x$ ranges from 1 to $\CardX$.
\EDef

As an alternative for the designation \POVM, one sometimes sees the 
term ``Probability Operator Measure'' used in the literature.
It is a postulate of quantum mechanics that any physically 
realizable measurement can be described by a \POVM.  Moreover, for
every \POVM, there is {\em in principle\/} a physical procedure with
which to carry out the associated measurement.  Therefore, we can
denote the set of all possible measurements, or equivalently the
set of all \POVM{}s, as $\AllMmnts$.

\Warning
It should be noted that the scheme of measurements defined here
is the most general that can be contemplated within quantum
mechanics.  This is a convention that has gained wide usage within
the physics community only relatively recently (within the last 15
years or so).  Indeed, almost all older textbooks on quantum
mechanics describe a more restrictive notion of measurement.
In the usual approach, as developed by von Neumann, measurements are
taken be in one-to-one correspondence with the set of all Hermitian
operators on the given Hilbert space.  The eigenvalues of these
operators correspond to the possible measurement results.
The framework of POVMs described
above can be fit within the older von Neumann picture if one is
willing to take into account a more detailed picture of the
measurement process, including all ancillary devices used along the
way.  The ultimate equivalence of these two pictures is captured
by a formal result known as Neumark's Theorem \MyCit{PeresBook}.

A Projection Valued Measurement (\PVM)---another name for 
the von Neumann measurements just described---is a special case of
a \POVM:  it is given
by adding the requirement that $E_x E_y = \delta(x,y) E_x$
(with $\delta(x,y) \Is 1$ if $x \Is y$ and 0 otherwise---%
\ie the Kronecker-delta).  With this requirement,
the operators $E_x$ are necessarily projection operators, and
so can be thought of as the eigenprojectors of an Hermitian operator.
One consequence of this is that the number of outcomes in a PVM can
never exceed the dimensionality of the Hilbert space.  General POVMs
need not be restricted in this way at all; moreover the $E_x$ need
not even commute.

\BEx
Measuring whether a photon is polarized according
to angle $\alpha$ or to $\alpha + \pi/2$ is done
by the \POVM
\[
\left\LAcc \Matrix{c^2 & cs}{cs & s^2}, 
\Matrix{s^2 & -cs}{-cs & c^2}\right\RAcc,
\]
where $c = \cos\alpha$ and $s = \sin\alpha$.
This is a \PVM.
When applied to a photon
known to be in state \Ket{\Hor}, for instance,
this results in the \aaPD 
$\LA c^2, s^2 \RA$,
using equation \RefEq{TraceStateOp}.

An example of a \POVM which is not a \PVM is the
symmetric three-outcome ``trine'' \POVM:
let $\gamma = \cos(\pi/3)$ and
$\sigma = \sin(\pi/3)$
\[
\left\LAcc \TwoThird \Matrix{1 & 0}{0 & 0},
\TwoThird \Matrix{\gamma^2 & \gamma\sigma}%
                 {\gamma\sigma & \sigma^2},
\TwoThird \Matrix{\gamma^2 & -\gamma\sigma}%
                 {-\gamma\sigma & \sigma^2}
\right\RAcc,
\]
which simplifies to
\[
\left\LAcc \Matrix{ \TwoThird & 0}{0 & 0},
\Matrix{\OneSixth & \OneSixth \sqrt{3}}%
                 {\OneSixth \sqrt{3} & \Half},
\Matrix{\OneSixth & -\OneSixth \sqrt{3}}%
                 {-\OneSixth \sqrt{3} & \Half}
\right\RAcc.
\]
Applying this \POVM to the state $\Ket{\Vert}$
results in the \aaPD
$\LA 0, \Half, \Half \RA$,
again according to \RefEq{TraceStateOp}.
\EEx

There are two advantages to using the formalism of POVMs over that
of PVMs.  First, it provides a compact formalism for describing
measurements that the PVM formalism has to stretch to obtain---by
considering ancillary systems, extra time evolutions, etc., in the
measurement process.  Second, and most importantly, there are some
situations that call for all these extra steps to obtain an optimal
measurement.  A simple example is that of having to distinguish
between three possible states for a system with a
two-dimensional Hilbert space:  the optimal POVM will generally have
three outcomes, whereas a direct von Neumann measurement on the
system can only have two.

%%%%%%%%%%%%%%%%%%%%%%%%%%%%%%%%%%%%%%%%%%%%%%%%%%%%%%%%%%%%
\section{Measures of \aaDisty}
%% CLASSICALLY

We have just seen that a measurement (a \POVM) applied to a \aaDM
results in a \aaPD. Suppose now we have two \aaDMs defined
over the same Hilbert space.  Then we find ourselves back in the
(classical) situation described in the previous section: comparing
two \aaPDs over the same outcome space $X$. In particular,
let $\rho_0$ and $\rho_1$ be two density matrices,
and let $\CalE = \{ E_1, \ldots  , E_\CardX \}$ denote a \POVM.
Let $\PZE$ denote the probability \aaDistn obtained by performing
the \POVM \CalE on a system in state $\rho_0$ according to equation
\RefEq{TraceStateOp}; let \PIE be defined likewise.
Then we have:

\BCenter
\begin{tabular}{| l | l || l l l l l |}
\hline \NoMathAfterHLS \BeforeHLS
          & $$   & $x \Eq 1$        & $x \Eq 2$
& $
x \Eq 3$        & \ldots  &  $x \Eq \CardX$  \\ \hline \hline
 
\MathAfterHLS
\PZE & $\pi_0$ & $\Trace(\rho_0 E_1)$ & $\Trace(\rho_0 E_2)$ &
$\Trace(\rho_0 E_3)$ & \ldots  & $\Trace(\rho_0 E_\CardX)$ \\ %
 
\MathBetweenRowsS \BeforeHLS
\PIE & $\pi_1$ & $\Trace(\rho_1 E_1)$ & $\Trace(\Rho{1} E_2)$ &
$\Trace(\rho_1 E_3)$ & \ldots  & $\Trace(\Rho{1} E_\CardX)$ 
\\ %
\hline
\end{tabular}\newline
\ECenter
As before, $\pi_0$ and $\pi_1$ denote the {\em a priori\/}
probabilities and are assumed to be equal to \Half.

This section discusses four notions of distinguishability for
probability distributions and---by way of the connection above---also
density matrices.  The unique feature in the quantum case is
given by the observer's freedom to choose the measurement.  Since, of
course, one would like to choose the quantum measurement to be as
useful as possible, one should optimize each distinguishability
measure over all measurements:  the values singled out by this
process gives rise to what we call the quantum distinguishability
measures.

The reader should note that being able to distinguish between
probability distributions---that is, between alternative statistical
hypotheses---is already an important and well-studied problem
with a vast literature.  It goes under the name of statistical
classification, discrimination or feature evaluation, and has had
applications as far-flung as speech recognition and radar detection.
For a general overview, consult \MyCit{BenBassat82}.  The problem
studied here is a special case of the general one, in the sense that
we want to distinguish between two (and only two) discrete \aaPDs
with equal a priori \aaProbs.

In the following subsections each classical measure of
distinguishability is discussed first, followed by a discussion of
its quantum counterpart.

\subsection{Probability of error}
\label{CoffeeGallery}

Consider the following experimental situation where \B is asked to
distinguish between two stochastic variables.  \A provides him with
one sample, $x$, with equal probability to have been secretly chosen
from either $\SVX_0$ or $\SVX_1$.  \B's task is to guess which of the
two stochastic variables the sample came from, $\SVX_0$ or $\SVX_1$.
Clearly, the average probability that \B  makes the right guess
serves as a measure of \aaDisty between the two probability
distributions $p_0(x)$ and $p_1(x)$.

It is well known that \B's optimal strategy is to look at the
\emph{a posteriori} probabilities:  given the sample $x$, his best
choice is the $t$ for which $r_t(x) $ is maximal (see the
representation at the end of Section \MyRef{Sec:PDs}). This strategy
is known as \emph{Bayes' strategy}.  So the average probability of
successfully identifying the distribution equals
$\sum_{x \in X} p(x) \max\{ r_0(x) , r_1(x) \}=$
$\Half  \sum_{x \in X} \max\{ p_0(x) , p_1(x) \}$.
Conversely, we can also express the \aaPry that \B fails.

\BDef
The \DefEmph{probability of error between two
\aaPDs{}} is defined by
\BEqn
\CPE (p_0, p_1) \IsDef \Half  \sum_{x \in X} \min\{p_0(x),p_1(x)\}
\EEqn
Two identical \aaDistns have $\CPE = \Half$,
and two \aaOrth \aaDistns have $\CPE = 0$.
\EDef

\Warning
\CPE is not a distance function:  for example, when two \aaDistrns
are close to one another, \CPE is \emph{not} close to $0$,
but close to \Half.

%% QUANTUM

In the quantum-mechanical case, the experimental set-up is almost
identical.  \A has two ensembles, one according to $\rho_0$, the
other according to $\rho_1$.  She provides \B with a quantum
sample chosen from one of the two ensembles with equal \aaPry.
Following a measurement, \B must again guess from which ensemble the
sample was drawn: the one under $\rho_0$ or the one under $\rho_1$.

For any fixed measurement, the Bayesian strategy of guessing
the density operator with the largest posterior probability is the
optimal thing to do.  However, now \B should as well make use of his
extra degree of freedom:  he can choose the measurement he applies to
his sample.  He should choose the \aaMmnt that minimizes his \aaPOE.
So we define:

\BDef
The \DefEmph{probability of error between two
density matrices} $\rho_0$ and $\rho_1$ is defined by
\BEqn
\QPE(\rho_0,\rho_1) \IsDef
\min_{\CalE \in \AllMmnts} \CPE(p_0(\CalE),p_1(\CalE)),
\EEqn
where  the \POVM \CalE ranges over the set of all
possible measurements \AllMmnts.
\EDef

(More carefully, one should use ``infimum'' in this definition.
However---since in all the optimization problems we shall consider
here, the optima actually can be obtained---there is no need for the
extra rigor.)

The question of finding an explicit formula for the optimal \POVM
in this definition was first studied by
Helstrom~\MyCitt{HelstromBook}{pp. 106--108}.
He shows that the \POVM $\CalE^{\ast}$ that \emph{minimises}
$\CPE(p_0(\CalE),p_1(\CalE))$
is actually a \PVM.
Knowing the optimal \POVM, the \aaPOE can be expressed explicitly.
The expression he gives is,
\BEqn
\label{Burp}
\QPE(\rho_0,\rho_1) 
   = \Half + \Half \sum_{\lambda_j\le0}  \lambda_j\;,
\EEqn
where the ${\lambda_j}$ denote the eigenvalues of the
matrix $\Gamma= \rho_0 - \rho_1 $.

This expression can be cleaned up a little in the following way.
Consider the function $f(x)=\Half (x-|x|)$.  It vanishes when
$x\ge0$ and is the identity function otherwise.  Thus, with its
use, we can expand the summation in Eq.~(\ref{Burp}) to be over
all the eigenvalues of $\Gamma$:
\BEqnA
\QPE(\rho_0,\rho_1)
& = &
\Half +  \Half \sum_{j=1}^N f( \lambda_j )\\
& = &
\Half + \Quart\Trace\Gamma - \Quart\sum_{j=1}^N | \lambda_j |\\
&=&
\Half - \Quart\Trace|\Gamma|\;.
\EEqnA

Hence we have the following proposition:
\begin{Proposition}
Given two arbitrary \aaDMs  $\rho_0$ and $\rho_1$,
the \aaPOE equals

\BEqn \LabelEq{QPEIsTN}
\QPE(\rho_0,\rho_1) 
   = \Half - \Quart \sum_{j=1}^N | \lambda_j |
   = \Half - \Quart \Trace | \rho_0 - \rho_1 |,
\EEqn
where the ${\lambda_j}$ are the eigenvalues
of $ \rho_0 - \rho_1 $.
\EProp

$\QPE(\rho_0,\rho_1)$ is therefore just a simple function of the
distance between $\rho_0$ and $\rho_1$, when measured as the trace
norm of their difference.  (An alternative derivation of this
can be found in \MyCit{Fuchs96}.)

%%%%%%%%%%%%%%%%%%%%%%%%%%%%%%%%%%%%%%%%%%%%%%%%%%%%%%%%%%%%
\subsection{Kolmogorov distance}

Among (computational) cryptographers, another measure of \aaDisty
between probability distributions is used fairly often: the standard
notions of exponential and computational indistinguishability
\MyCit{Yao82b, GoldwasserMi84, GoldwasserMiRa89} are based on it.

\BDef
The \DefEmph{Kolmogorov distance between two \aaPDs{}}
%%  (also called \emph{variational distance})
is defined by
\BEqn
\CK(p_0, p_1) \IsDef \Half \sum_{x \in X} | p_0(x) - p_1(x) |.
\EEqn
Two identical \aaDistns have $\CK = 0$,
and two \aaOrth \aaDistns have $\CK = 1$.
\EDef

In some references the factor of \Half plays no role, and the
``Kolmogorov distance'' is defined without it.  Here we have included
it because we wanted $\CK$ to take values between \Zero and \One.

%%  The operational definition of $\CK$ is as follows:

Probability of error and Kolmogorov distance
are closely related.

\BProp
\BEqn \LabelEq{PEIsK}
\CPE (p_0, p_1) = \Half - \Half \CK(p_0, p_1)
\EEqn
\EProp

This is not very difficult to prove.  The most important step is to
split the sum over $X$ into two disjoint sub-sums, one for which
$p_0(x) < p_1(x)$, and one for which $p_0(x) \geq p_1(x)$.
See \MyCit{Toussaint72}.

%% QUANTUM
In the quantum case, we must again optimise over all possible
measurements.  But here this means that we have to find the \POVM
that maximises the \aaKD.
\BDef
The \DefEmph{\aaKD between two
density matrices} $\rho_0$ and $\rho_1$ is defined by
\BEqn
\QK(\rho_0,\rho_1) \IsDef
\max_{\CalE \in \AllMmnts} \CK(p_0(\CalE),p_1(\CalE)),
\EEqn
where  the \POVM \CalE ranges over the set of all
possible measurements \AllMmnts.
\EDef

The relation between \aaPOE and \aaKD (eq. \MyRef{Eq:PEIsK}) shows
that the two \aaMmnt{}s that optimise \CPE and \CK are identical:
\EOpt minimises the function
$\CPE(\PZE, \PIE)$
\aaIff it also
maximises $\CK(\PZE,\PIE)$.
See also the appendix of \MyCit{Fuchs96}.
Combining equations \RefEq{QPEIsTN} and \RefEq{PEIsK} we get:

\BProp
The \aaKD between two \aaDMs $\rho_0$ and $\rho_1$ equals
\BEqn
\QK(\rho_0,\rho_1)
= \Half  \sum_{j=1}^N | \lambda_j |
= \Half \Trace | \rho_0 - \Rho{1} |,
\EEqn
where the ${\lambda_j}$ are the eigenvalues
of $\rho_0 - \rho_1$.
\EProp

Observe that $\Trace | \rho_0 - \rho_1 |$ is simply the
\emph{trace-norm distance} on operators
\MyCit{ReedSimonBook,Orlowski96}.  Hence \QK has the additional
property of satisfying a triangle inequality.  The trace-norm
distance appears to be of unique significance within the class of all
operator-norms because of its connection to probability of error.

%%%%%%%%%%%%%%%%%%%%%%%%%%%%%%%%%%%%%%%%%%%%%%%%%%%%%%%%%%%%%%%%%%%%%
\subsection{Bhattacharyya coefficient}

Another distinguishability measure that has met widespread use---%
mostly because it is sometimes easier to evaluate than the
others---is the Bhattacharyya coefficient.  See
\MyCit{Kailath67, Toussaint71, Toussaint72}.

\BDef
The \DefEmph{Bhattacharyya coefficient between two \aaPDs{}}
 $p_0$ and $p_1$ is defined by
\BEqn
\CB(p_0, p_1) = \sum_{x \in X} \sqrt{ p_0(x)  p_1(x)}\;.
\EEqn
Two identical \aaDistns have $\CB = 1$,
and two \aaOrth \aaDistns have $\CB = 0$.
\EDef

\Warning
\CB is also not a distance function:  for instance, when two
\aaDistrns are close to one another, \CB is \emph{not} close to $0$.
It can however be easily related to a distance function by taking
its arccosine.

The Bhattacharyya coefficient's greatest appeal is its
simplicity:  it is a sort of overlap measure between the two
distributions.  When their overlap is zero, they are completely
distinguishable; when their overlap is one, the distributions are
identical and hence indistinguishable.  Moreover the Bhattcharyya
coefficient can be thought of geometrically as an inner product
between $p_0$ and $p_1$, interpreted as vectors in an
$m$-dimensional vector space.  However, it does not appear to bear a
simple relation to the probability of error in any type of
statistical inference problem.

%% QUANTUM

In the quantum case, we define a distinguishability measure
by minimising over all possible measurements.

\BDef
The \DefEmph{\aaBhCo between two
density matrices} $\rho_0$ and $\rho_1$ is defined by
\BEqn
\QB(\rho_0,\rho_1) \IsDef \min_{\CalE \in \AllMmnts}
                            \CB(p_0(\CalE),p_1(\CalE)),
\EEqn
where the \POVM \CalE ranges over the set of all
possible measurements \AllMmnts.
\EDef

The following proposition provides a closed-form expression for this
distinguishability measure.
\BProp \MyLabel{Th:OverlapIsQB}
\ThAttrTo{(Fuchs and Caves~\MyCit{FuchsCa95})}
The quantum Bhattacharyya coefficient can be expressed as
\BEqn
\label{Snort}
\QB(\rho_0,\Rho{1})=
 \Trace\left(\sqrt{\sqrt{\rho_0} \Rho{1} \sqrt{\rho_0}} \right),
\EEqn
where the square-root of a matrix $\rho$ denotes any positive
semi-definite matrix $\sigma$ such that $\sigma^2=\rho$.
\EProp

Surprisingly, it turns out that \QB is equivalent to another
{\em non-measurement oriented\/} notion of distinguishability.
Suppose \KPsi{0} and \KPsi{1} are pure states.  When
we think these two state vectors geometrically, a natural notion of
distinguishability is the angle between \KPsi{0} and \KPsi{1}, or any
simple function of this angle like the inner product or overlap.  In
particular, we can define $ overlap( \Ket{\psi_0}, \Ket{\psi_1} ) 
:= \Abs{\Inner{\psi_0}{\psi_1}}$ as a measure of distinguishability.
The question is: what to do for mixed states?

The answer was given by Uhlmann~\MyCit{Uhlmann76}\footnote{A nice
review of this theorem in terms of finite-dimensional Hilbert space
methods can be found in \MyCit{Jozsa94}.}.  If $\rho_0$ is the \aaDM
of a mixed state in the Hilbert space $\CalH_1$, then we can always
extend the
Hilbert space such that $\rho_0$ becomes a pure state in the combined
Hilbert space $\CalH_1 \otimes \CalH_2$.  More precisely, we can
always find an extension $\CalH_2$ of $\CalH_1$ and a pure state
$\Ket{\psi_0} \in \CalH_1 \otimes \CalH_2$, such that 
$\Trace_2 ( \Proj{\psi_0} ) = \rho_0 $.  Here the operator
$\Trace_2 $ means to perform a partial-trace operation over the
ancillary Hilbert space $\CalH_2$.
When this condition holds, \Ket{\psi_0} is said to be a
{\em purification} of $\rho_0$.  Similarly, if \KPsi{1} is the
purification of $\rho_1$, we are back to a situation with two pure
states, and we can apply the formula above, leading to the following
generalised definition.
\BDef
The \DefEmph{(generalised) overlap} between two \aaDMs is defined by
\BEqn \LabelEq{Overlap}
\Overlap(\rho_0,\rho_1) \IsDef \max \Abs{\Inner{\phi_0}{\phi_1}},
\EEqn
where the maximum is taken over all purifications
$\Ket{\phi_0}$ and $\Ket{\phi_1}$
of $\rho_0$ and $\rho_1$ respectively.
\EDef
It can be demonstrated that \MyCit{FuchsCa95},
\BEqn \LabelEq{USRE}
overlap(\rho_0,\rho_1) = \QB(\rho_0,\Rho{1}).
\EEqn

Despite the rather baroque appearance $\QB(\rho_0,\Rho{1})$
takes in Eq.~(\ref{Snort}), it is endowed with several very nice
properties.  For instance, $\QB(\rho_0,\rho_1)$ 
is multiplicative over tensor products:
\BEqn
\QB(\rho_0 \otimes \rho_1, \rho_2 \otimes \rho_3) = 
\QB(\rho_0, \rho_2) \; \QB(\rho_1, \rho_3).
\EEqn
$\QB$'s square is concave over one of its arguments; i.e.,
if $0 \leq \mu_0,\mu_1 \leq 1, \mu_0 + \mu_1 \Is 1$ then
\BEqn
\Big(\QB(\rho, \mu_0 \rho_0 + \mu_1 \rho_1)\Big)^2 \geq
\mu_0\Big(\QB(\rho,  \rho_0)\Big)^2 +
\mu_1 \Big(\QB(\rho,  \rho_1)\Big)^2.
\EEqn
Moreover, $\QB$ itself is doubly concave\footnote{We thank C.~M.
Caves for pointing this out to us.}:
\BEqn
\QB(\mu_0 \rho_0 + \mu_1 \rho_1, \mu_0 \rho_2 + \mu_1 \rho_3)
\geq
\mu_0\QB(\rho_0,  \rho_2) +
\mu_1\QB(\rho_1,  \rho_3).
\EEqn

%%%%%%%%%%%%%%%%%%%%%%%%%%%%%%%%%%%%%%%%%%%%%%%%%%%%%%%%%%%%%%%%%%%%
\subsection{\aaSD}

Now we come to the last, and maybe most important, notion of
distinguishability.  Mutual information, as defined by Shannon
\MyCit{Shannon48}, can be used as a \aaDisty measure between
probability distributions \MyCit{Lindley56,BenBassat82}.  We assume
that the reader is familiar with the (Shannon) entropy function
$\ShE$, the argument of which can be either a \aaSV or a \aaPD.
$\BShE(p)=-p\log p - (1-p)\log(1-p)$ is the entropy of the
distribution $\LA p, 1-p \RA$.

Consider the following elementary example.  Suppose we have two
boxes, each containing colored balls.  Let $t \in T \Is \ZO$ denote
the identity of the boxes; and let us think of \SVT as a stochastic
variable. Then $ \Prob[ \SVT=t ] $ is just the
a priori \aaProby $\pi_t$ of Section \MyRef{Sec:PDs}.
Recall that in our case $\pi_0 = \pi_1 = \Half$, so $\ShE(\SVT) = 1$.
Let \SVX denote the stochastic variable corresponding
to the color of a ball upon being drawn from a box, taking into
account that the identity of the box is itself a stochastic variable.
Recall that $\Prob[\SVX \Is x]$ was written as $p(x)$.

Consider the same experiment as in Subsection~\ref{CoffeeGallery},
in which \A picks a ball from one of the two boxes and gives
it to \B.  One can ask now:  How much information does \SVX (the
color of a picked ball) convey about \SVT (the identity of the box it
came from)?

Information is defined as the reduction of uncertainty, where
uncertainty is quantified using the Shannon entropy.  Consider two
quantities: \ili1 the average uncertainty of \B about \SVX before he
was handed a sample (or ball); and \ili2 his average uncertainty
about \SVX after  he was handed a sample.  This difference expresses
the amount of information gained through the experiment.  It can also
be used as a measure of distinguishability between two distributions.
When there is no difference between the distributions, the amount of
information that can be gained in this way is zero.  When the
distributions are orthogonal, all the information about \SVT can be
gathered.  Thus we obtain:
\BEqnA
{\rm average\ information}& = & ({\rm average\ uncertainty\ about\ }
\SVX) - \\
&   & \;\quad ({\rm average\ uncertainty\ about\ } \SVX
%%         {\rm \ when\ } t {\rm \ is \ given)} \\
       {\rm \ given\ } t ) \\
& = & \ShE(p(x)) - (\Half \ShE(p_0(x)) + \Half \ShE(p_1(x))) \\
& = & \ShE(\SVX) - \sum_{t \in T} \Prob [\SVT=t] \; \ShE(\SVX | \SVT = t) \\
& = & \ShE(\SVX) - \ShE(\SVX | \SVT) \\
& = & I(\SVT ; \SVX).
\EEqnA

This leads to the following definition:
\BDef \MyLabel{DefSD}
The \DefEmph{\aaSD between two \aaPDs{}}
$p_0$ and $p_1$ is defined by:
\BEqn
\CSD (p_0, p_1) \IsDef  \ShI(\SVT ; \SVX).
\EEqn
\EDef

Since the mutual information is symmetric in its two random
variables, it can also be expanded in the other direction to look
like:
\BEqnA
\CSD (p_0, p_1) & = & \ShI(\SVT, \SVX) \\
& = & \ShE(\SVT) - \ShE(\SVT | \SVX) \\
& = & 1 - \sum_{x \in \SetX} p(x) \; \ShE(\SVT | \SVX = x ) \\
& = & 1 - \sum_{x \in \SetX} p(x) \; \BShE(r_0(x))
\EEqnA
This form will be useful for various of the proofs to come.

In the same fashion as all the other distinguishability measures,
the \aaSD can be applied to the quantum case.  We must find the
measurement that optimises it when tabulated for probability
distributions obtained by applying a quantum measurement.
\BDef \MyLabel{DefQSD}
The \DefEmph{\aaSD between two density matrices}
$\rho_0$ and $\rho_1$ is defined as
\BEqn
\QSD(\rho_0,\rho_1) \IsDef \max_{\CalE \in \AllMmnts}
                            \CSD(p_0(\CalE),p_1(\CalE)),
\EEqn
where  the \POVM \CalE ranges over the set of all
possible measurements \AllMmnts.
\EDef

There is an unfortunate problem for this measure of distinguishabity:
calculating the value $\QSD(\rho_0,\rho_1)$ is generally a difficult
problem. Apart from a few special cases, no explicit formula for
$\QSD$ solely in terms of $\rho_0$ and $\rho_1$ is known.  Even
stronger than that: no such formula can exist in the general case
\MyCit{FuchsCa94}.  This follows from the fact that optimising the
Shannon distinguishability requires the solution of a transcendental
equation.  (See also \MyCit{PeresBook} and \MyCit{MorPhD} for
a discussion of other aspects of $\QSD$.)

%%%%%%%%%%%%%%%%%%%%%%%%%%%%%%%%%%%%%%%%%%%%%%%%%%%%%%%%%%%%%%%%%%%%
\subsection{Overview}

The material presented in the previous four subsections
can be summarized in the following table. \\[1ex]
{\small
\begin{tabular}{|l|lllll|}
\hline
\NoMathAfterHLS
& classical &  when  & when  & optimality & quantum \\ 

\BeforeHLS
& definition & $p_0 = p_1$ & $p_0 \Orth p_1$ & criterion & expression
\\ \hline 

\MathAfterHLS
{\sc pe} & $ \Half \sum \min\{ p_0(x) , p_1(x) \} $&
1/2 & 0 & min & $\Half - \Quart \Trace | \rho_0 - \rho_1 |$\\[4mm]

\MathBetweenRowsS
{\sc k}  & $ \Half \sum | p_0(x) - p_1(x) | $ &
0 & 1 & max & $ \Half  \Trace |  \rho_0 - \rho_1  |$ \\[4mm]

\MathBetweenRowsS
{\sc b} & $ \sum \sqrt{p_0(x)p_1(x)} $ &
1 & 0 & min & $\Trace\sqrt{\rho_0^{1/2}\rho_1\rho_0^{1/2}}$ \\[4mm]

\MathBetweenRowsS
{\sc sd} & $ \textit{via\ } I(\SVX ; \SVT) $ &
0 & 1 & max & \textrm{no simple form} \\[4mm] \hline

\end{tabular}
\newline
}

\section{Inequalities}

We have seen already (Eqn.~\RefEq{PEIsK}) that \aaPE and \aaKD
are related through the equality:
%%  \CPE vs. \CK:
\BEqn
\CPE (p_0, p_1) = \Half - \Half \CK(p_0, p_1)
%%  \LabelEq{PEIsK}
\EEqn
The other pairs of \aaDisty measures are related through
inequalities, some of which can be found in the literature
\MyCit{Kailath67, HellmanRa70, Toussaint71, Toussaint72,%
BenBassat82}.
\BProp \MyLabel{Th:Ineqs}
%%  \CPE vs. \CB
Let $p_0$ and $p_1$ be \aaPDs.  The following relations hold:
%%  
%%  \CPE vs. \CK
\BEqn
 \LabelEq{BLeqK}
1 - \CB(p_0, p_1) \;\;\leq\;\;
 \CK(p_0, p_1) \;\;\leq\;\;
\sqrt{1-\CB(p_0, p_1)^2}\;,
\EEqn
%%  
%%  \CPE vs \CSD
\BEqn
\LabelEq{PELeqSD}
1 - \BShE\Big(\CPE (p_0, p_1)\Big) \;\;\leq\;\; \CSD (p_0, p_1)
\;\;\leq\;\; 1 - 2 \CPE (p_0, p_1)\;,
\EEqn
%%  
%%  \CB vs \CSD
\BEqn \LabelEq{BLeqSD}
1 -  \CB (p_0, p_1)
\;\;\leq\;\; \CSD (p_0, p_1) \;\;\leq\;\;
1-\BShE\!\left(\Half-\Half\sqrt{1-\CB (p_0, p_1)}\right).
\EEqn
\EProp

Before giving the proof of this proposition, we state its quantum
equivalent.  This is the main result of the paper.

\BTh \MyLabel{Th:QIneqs}
Proposition \MyRef{Th:Ineqs} can be generalized to the quantum
scenario: one can substitute \QPE, \QK, \QB and \QSD
and use \aaDMs  $\rho_0$ and $\rho_1$ as operands.
Alternatively, using the quantum expressions,
equations \RefEq{BLeqK}, \RefEq{PELeqSD} and  \RefEq{BLeqSD}
can be expressed in the following, equivalent form:
\BEqn \LabelEq{QBLeqK}
1 - \QB(\rho_0,\rho_1) \;\;\leq\;\;
 \Half  \Trace |  \rho_0 - \rho_1|  \;\;\leq\;\;
\sqrt{1-\QB(\rho_0,\rho_1)^2}\;,
\EEqn
%% 
%%  \CPE vs \CSD
\BEqn \LabelEq{QPELeqSD}
1 - \BShE\Big(\Half - \Quart \Trace | \rho_0 - \rho_1 |\Big)
\;\;\leq\;\; \CSD (\rho_0, \rho_1)
\;\;\leq\;\; \Half  \Trace |  \rho_0 - \rho_1|\;, 
\EEqn
%% 
%%  \CB vs \CSD
\BEqn \LabelEq{QBLeqSD}
1 - \QB(\rho_0,\rho_1)
\;\;\leq\;\; \CSD (\rho_0, \rho_1)
\;\;\leq\;\;
1-\BShE\!\left(\Half-\Half\sqrt{1-\QB(\rho_0,\rho_1)}\right).
\EEqn
\ETh

The importance of this theorem is that, while the quantum Shannon
distinguishability is impossible to calculate in a closed form,
the inequalities provide a useful way to bound it.  We will use these
bounds in an application in the next section.

\textsc{Proof of proposition \MyRef{Th:Ineqs}:}\\
We start by proving Eq.~\RefEq{BLeqK}.
To get the left-hand inequality, note:
\BEqnA
1 -  \CB (p_0, p_1)
& = & \Half \left( \SumX p_0(x) + \SumX p_1(x) - 2 \;
\SumX \Sqrt{p_0(x) p_1(x)}\right) \\
& = & \Half \SumX
\left| \Sqrt{p_0(x)} - \Sqrt{p_1(x)} \right|^2 \\
& \leq & \Half  \SumX \Abs{ p_0(x) - p_1(x) } \\
& = & \CK (p_0, p_1)
\EEqnA
The inequality in the penultimate step holds for each term
individually.  To get the right-hand inequality, we simply use the
Schwarz inequality:
\BEqnA
\CK (p_0, p_1)^2
&=&
\Quart\!\left(\SumX \Abs{ p_0(x) - p_1(x) }\right)^2 \\
&=&
\Quart\!\left(\SumX\left|\Sqrt{p_0(x)} - \Sqrt{p_1(x)}\right|
\left| \Sqrt{p_0(x)} + \Sqrt{p_1(x)}\right|\right)^2  \\
&\leq&
\Quart\SumX\left( \Sqrt{p_0(x)} - \Sqrt{p_1(x)}\right)^2
\SumX\left(\Sqrt{p_0(x)} + \Sqrt{p_1(x)}\right)^2\\
%&=&
%\Quart\!\left(\SumX p_0(x)+p_1(x)-2\Sqrt{p_0(x)p_1(x)}\right)\!
%\!\left( \SumX p_0(x) + p_1(x) + 2 \Sqrt{p_0(x) p_1(x)}\right) \\
&=&
\Quart\Big(2 - 2 \CB (p_0, p_1)\Big)\Big(2 + 2 \CB (p_0, p_1)\Big)\\
&=&
1 - \CB (p_0, p_1)^2
\EEqnA

In order to prove the left inequality of Eq.~\RefEq{PELeqSD},
we observe that this is a special case of the 
Fano inequality (see for instance \MyCit{CoverTh91}):
\[
\ShE(\SVT|\SVX) \leq \BShE( \CPE (p_0, p_1) ) + 
\CPE (p_0, p_1)\;\log(\Hash T-1).
\]
where $\Hash T = 2$ is the cardinality of the set $T$.

For the right-hand inequality of Eq.~\RefEq{PELeqSD} we expand
$\ShI(\SVX ; \SVT)$  as $ \ShE(\SVT) - \ShE(\SVT | \SVX)$ to obtain
an inequality between $\CSD$ and $\CPE$.  (See also
\MyCit{HellmanRa70}.)  Recall the definitions of $r_t(x)$ and $p(x)$,
observing that $r_1(x) = 1 - r_0(x)$ and that
$2 \min \{ r, 1-r \} \leq \BShE(r)$ for all $r$ between $0$ and $1$
(see Figure \MyRef{fig:qind1}).  Hence, we obtain:
\BEqnA
\CSD (p_0, p_1) & = & 1 - \sum_{x \in \SetX} p(x) \; \BShE(r_0(x)) \\
& \leq & 1 - \sum_{x \in \SetX} p(x) \cdot 
                 2 \min \{ r_0(x), 1-r_0(x) \} \\
& \leq & 1 - 2 \CPE (p_0, p_1).
\EEqnA

The left-hand inequality of Eq.~\RefEq{BLeqSD} is obtained in a
similar way.  Using the fact that
$\BShE(r) \leq  2 \Sqrt { r(1-r) }$ (see Figure \MyRef{fig:qind1}),
we get:
\BEqnA
\CSD (p_0, p_1) & = &  1 - \sum_{x \in \SetX} p(x) \; \BShE(r_0(x))\\
& \geq & 1 - \sum_{x \in \SetX} p(x) \cdot
                   2 \sqrt{ r_0(x) (1-r_0(x)) } \\
& = &  1 - \sum_{x \in \SetX} \sqrt{ p_0(x) p_1(x) } \\
& = &  1 - \CB(p_0, p_1).
\EEqnA

For the right-hand side of Eq.~\RefEq{BLeqSD},
we define the function
\BEqn
 g(r) := \Half - \Half \Sqrt{1-r^2}
\EEqn
It can be verified that $k(r) =  2 \Sqrt { r(1-r) }$ is the
inverse of $g(r)$ when $0 \le r \le \Half$.  Moreover, 
$k(r)=k(1-r)=\min\{ r, 1-r)\}$.  Using the fact that
$\BShE(r)=\BShE(1-r)=\BShE(\min \{ r, 1-r) \})$ and 
that $\BShE(g(r))$ is a convex function, we obtain by Jensen's
inequality that
\BEqnA
\CSD (p_0, p_1) & = &  1 - \SumX p(x) \; \BShE(r_0(x)) \\
& = &1-\SumX p(x)\;\BShE\Big(\!\min \{ r_0(x), 1-r_0(x) \}\Big)\\
& = &1-\SumX p(x)\;\BShE\Big(g\Big(k\Big(\!\min\{r_0(x),1-r_0(x)\}
     \Big)\Big)\Big)\\
& = & 1 - \SumX p(x) \; \BShE\Big(g(\;k(r_0(x))\;)\Big)  \\
&\le&1-\BShE\!\left(g\!\left(\SumX \; p(x) k(r_0(x))\right)\right) \\
& = &  1 - \BShE\Big(g( \CB (p_0, p_1) )\Big) \\
& = &1-\BShE\left( \Half - \Half \Sqrt{1-( \CB (p_0, p_1))^2}\right).
\EEqnA
This concludes the proof of Proposition  \MyRef{Th:Ineqs}.%
\EProof 

\begin{figure}
\BCenter
\epsfig{file=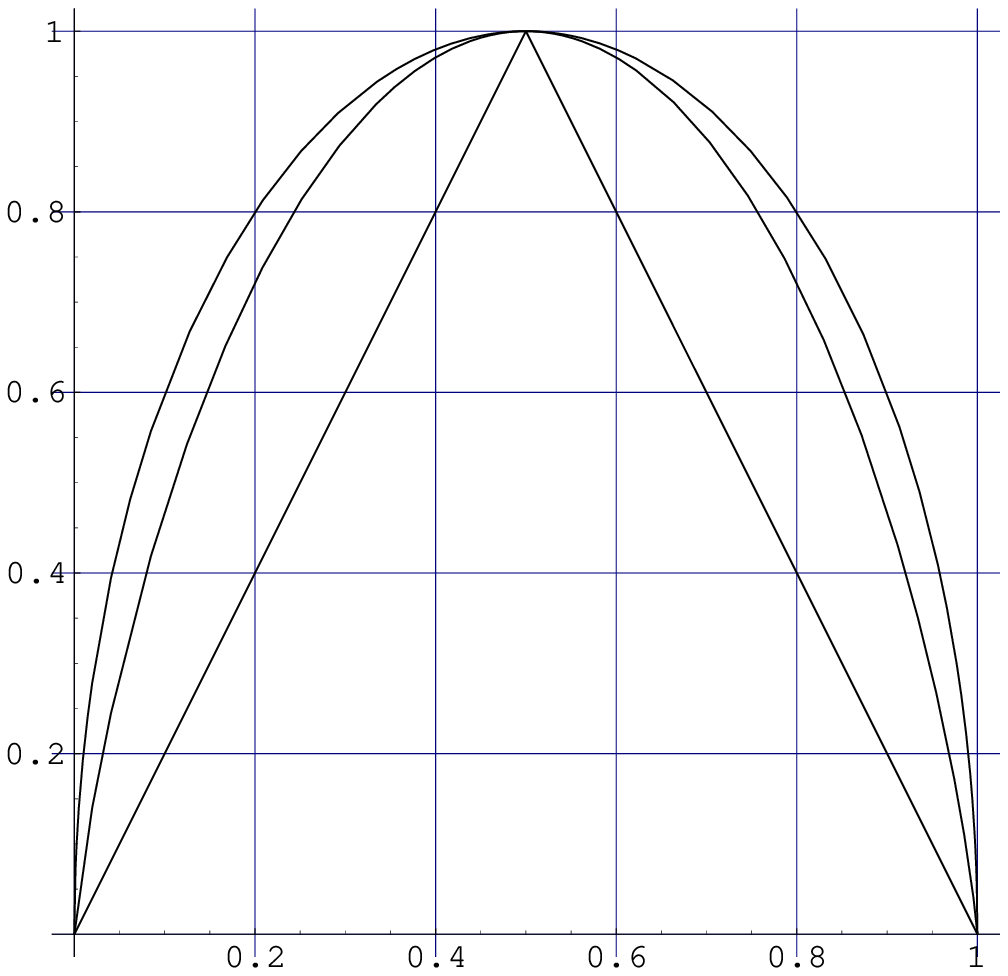,height=7cm}
\ECenter
\caption{\small
\ensuremath{2\min\LAcc x,1-x\RAcc\leq\BShE(x)\leq2(x(1-x))^{\Half}}
\protect\\ (Formally this is proven by looking at the first and
second derivatives.) }
\MyLabel{fig:qind1}
\end{figure}

The main tool in proving the quantum versions of these inequalities
is in noting that all the bounds are appropriately monotonic in
their arguments.

\textsc{Proof of theorem \MyRef{Th:QIneqs}:}\\
First we prove Eq.~\RefEq{QBLeqK}. We start with the first
inequality.  Let $\EOpt_{B}$ denote a \POVM that optimizes $\CB$ and
define $\EOpt_{K}$ likewise.
\BEqnA
1 - \QB(\rho_0,\rho_1) & = &
 1 - \CB\Big(p_0(\EOpt_{B}),p_1(\EOpt_{B})\Big) \\
  & \leq & \CK\Big(p_0(\EOpt_{B}),p_1(\EOpt_{B})\Big) \\
  & \leq & \CK\Big(p_0(\EOpt_{K}),p_1(\EOpt_{K})\Big) \\
  & =    & \QK(\rho_0,\rho_1)
\EEqnA
The second inequality of Eq.~\RefEq{QBLeqK} follows from
\BEqnA
\QK(\rho_0,\rho_1) & = & \CK(p_0(\EOpt_{K}),p_1(\EOpt_{K})) \\
  & \leq & \sqrt{1 - \CB\Big(p_0(\EOpt_{K}),p_1(\EOpt_{K})\Big)^2} \\
  & \leq & \sqrt{1 - \CB\Big(p_0(\EOpt_{B}),p_1(\EOpt_{B})\Big)^2} \\
  & =    & \sqrt{1 - \QB(\rho_0,\rho_1)^2}
\EEqnA
Equations \RefEq{QPELeqSD} and \RefEq{QBLeqSD} are proven in an
identical way. In particular, in Eq.~\RefEq{QBLeqK}
the functions on the extreme left, $f(x) = 1-x$, and on the extreme
 right, $f(x) = \Sqrt{1 - x^2}$ are both monotonically decreasing.
In addition, $\CB$ must be minimized whereas $\CK$ must be maximized.
The same is true for Eqs.~\RefEq{QPELeqSD} and \RefEq{QBLeqSD}.
\EProof

%%%%%%%%%%%%%%%%%%%%%%%%%%%%%%%%%%%%%%%%%%%%%%%%%%%%%%%%%%%%%%%%%%%
\section{Exponential \aaIndisty}
As already described in the Introduction, in the solution of various
cryptographic tasks, one often actually designs a whole \emph{family}
of protocols.  These are parameterized by a \emph{security
parameter}, $n$:  a number that might denote the length of some
string, the number of rounds, or the number of photons transmitted,
for instance.  Typically the design of a good protocol requires that
the probability of cheating for each participant vanishes
exponentially fast, \ie it goes as $O(2^{-n})$.  As an example, one
technique is to compare the protocol implementation (the family of
protocols) with the \emph{ideal protocol specification} and to prove
that these two become exponentially indistinguishable
\MyCit{GoldwasserMiRa89, Beaver91}.

\BDef
Let $\{\SVX_0\}=\LA\SVX_0^{(1)},\SVX_0^{(2)},\SVX_0^{(3)},\ldots\RA$
denote a \DefEmph{family} of stochastic variables with corresponding
distributions $\LA p_0^{(1)}, p_0^{(2)}, p_0^{(3)}, \ldots \RA$ .
Let $\{ \SVX_1 \}$ be defined similarly.
Then $\{ \SVX_0 \}$ and $\{ \SVX_1 \}$ are
\DefEmph{\aaEyI{}} if there exists an $n_0$ and an $\epsilon$
between \Zero and \One
such that
\[
\forall n \geq n_0 : \CK(p_0^{(n)}, p_1^{(n)}) \leq \epsilon^n
\]
\EDef

Examples of exponentially indistinguishable stochastic-variable
families can be constructed easily.  For instance, let $\SVX_0^n$ be
uniformly distributed over $\ZO^n$, the set of strings of length
$n$. That is to say, for each $x \in \ZO^n$, we have
$p_0^{(n)}(x) = 2^{-n}$.
This defines the family of uniform distributions over $\{ \SVX_0 \}$.
Let $\{ \SVX_1 \}$  be defined identically, {\em except\/} that
$p_1^{(n)}({\sf 0}^n) = 0$, while  $p_1^{(n)}({\sf 1}^n) = 2^{-n+1}$.
So for $\{ \SVX_1 \}$, ${\sf 0}^n$, the word with all zeroes, has
zero probability; while ${\sf 1}^n$, the word with all ones, has
double the probability it had in the uniform disttribution.
Clearly the two families $\{ \SVX_0 \}$ and $\{ \SVX_1 \}$
are \aaEyI.

The reader should be aware that in \emph{computational} \aaCrh{}y
more refined notions of \aaDisty have been defined
\MyCit{GoldwasserMiRa89}.  For {\em polynomial indistinguishability},
it is only required that the families converge as fast as $1/n^k$,
for some $k > 0$.  Though we will not argue it formally here, it is
not hard to see that the proof of Lemma \MyRef{Th:EIEquiv}
generalizes to apply to polynomially indistinguishable families.

Yet another refinement is \emph{computational \aaIndisty{}}.  For it,
a sample is given to a Turing machine (or a poly-size bounded family
of circuits), and we look at the Kolmogorov distance of the possible
outputs.  After maximizing over all Turing Machines, we say the
stochastic-variable families are computationally indistinguishable
if the distance between them converges to zero polynomially fast.
Computational indistinguishability has turned out to be extremely
powerful for defining notions as pseudo-random number generators
\MyCit{Yao82b} and zero knowledge protocols \MyCit{GoldwasserMiRa89}.
All these notions of protocol indistinguishability have that in
common if a distinguisher is given a sample and restricted to
polynomial-time calculations, then he will not be able to identify
the source of the sample.

Here we shall follow the computational-cryptographic tradition in
defining \aaEI via the Kolmogorov distance.  However, this choice is
in no way crucial: the next lemma shows that we could have taken any
of the four \aaDisty measures.  In other words $\CK$, $\CPE$, $\CB$
and $\CSD$ turn out to be equivalent when we require exponentially
fast convergence.\footnote{There is a small technicality here:
\aaIndiste \aaDistrns
have $\CPE = \Half$ and $\CB = 1$,
so \aaEI means convergence to those values,
instead of convergence to 0,
as is the case with \CK and \CSD.}

\begin{Lemma} \MyLabel{Th:EIEquiv}
Let $ \{ \SVX_0 \}$ and  $\{ \SVX_1 \}$ be two families of stochastic
variables that are \aaEyI with respect to \emph{one} of the \aaDisty
measures $\CK$, $\CPE$, $\CB$, $\CSD$. Then $\{ \SVX_0 \}$ and 
$\{ \SVX_1 \}$ are \aaEyI  with respect to \emph{each}
of $\CK$, $\CPE$, $\CB$, $\CSD$.
\end{Lemma}

\Proof
The equivalence between \aaEI for $\CPE$ and $\CK$ follows from
Eq.~\RefEq{PEIsK}. The other equivalences follow from
Eqs.~\RefEq{BLeqK} through \RefEq{BLeqSD}.  For instance, the proof
that  \aaEI for \CK implies \aaEI for \CB goes as follows. Suppose
\BEqn
[ \exists n_0, \epsilon ][\forall n \geq n_0 ]\ :\ \
\CK(p_0^{(n)}, p_1^{(n)}) \leq \epsilon^n .
\EEqn
Using the \LHS of Eq.~\RefEq{BLeqK}, it follows at once that
$\CB (p_0^{(n)}, p_1^{(n)})\ge 1 - \epsilon^n$.  It then follows from
the fact that $\CB (p_0^{(n)}, p_1^{(n)})$ is bounded above by
unity, that we obtain the desired exponential convergence.

The other implications are proven in a similar way.  As far as
expressions involving $\BShE(x)$ are concerned, it is sufficient to
recall (see fig. \MyRef{fig:qind1}) that 
\BEqn
2 \min \LAcc x, 1-x \RAcc \leq \BShE(x) \leq 2 \sqrt{x (1-x)}
\EEqn
This concludes the proof.~%
\EProof

The obvious next step is to define \aaEI for
\aaDMs, and to show that the choice of the \aaDisty measure
is immaterial.

\BDef
Let
$ \{ \rho_0^{(n)} \}$
$= \LA \rho_0^{(1)}, \rho_0^{(2)}, \rho_0^{(3)}, \ldots  \RA $
denote a \DefEmph{family} of \aaDMs
defined over the Hilbert space \HS.
Let $\{ \rho_1^{(n)} \}$ be defined similarly.
Then $\{ \rho_0^{(n)} \}$ and $\{ \rho_1^{(n)} \}$ are
\DefEmph{\aaEyI{}} if there exists an $n_0$ and an $\epsilon$
between \Zero and \One
such that
\[
\forall n \geq n_0 : \CK(\rho_0^{(n)}, \rho_1^{(n)}) \leq \epsilon^n
\]
\EDef

An example that makes use of this definition will be presented in the
next section.  However, first let us conclude with the quantum
analogue of Lemma \MyRef{Th:EIEquiv}.

\BTh
Let $ \{ \rho_0^{(n)} \}$ and  $\{ \rho_1^{(n)} \}$ be two families
of \aaDMs which are \aaEyI
with respect to \emph{one} of the \aaDisty measures
$\QK$, $\QPE$, $\QB$,  $\QSD$.
Then $ \{ \rho_0^{(n)} \}$ and  $\{ \rho_1^{(n)} \}$
are \aaEyI with respect to \emph{each}
of $\QK$, $\QPE$, $\QB$,  $\QSD$.
\ETh

\BProof
This follows immediately from Lemma \MyRef{Th:EIEquiv} and
Theorem \MyRef{Th:QIneqs}.~%
\EProof

%%%%%%%%%%%%%%%%%%%%%%%%%%%%%%%%%%%%%%%%%%%%%%%%%%%%%%%
\section{Applications}

Let us now look at an application of the quantum-exponential
indistinguishability idea.  In particular, we look at the problem of
the parity bit in quantum key distribution as studied in
\MyCit{BennettMoSm96} (henceforth called \BMS).
Let $|\psi_0\rangle = \Vec{\cos \alpha}{\sin \alpha}$
and $|\psi_1\rangle = \Vec{\cos \alpha}{-\sin \alpha}$,
and let $\rho_0$ and $\rho_1$ be the corresponding \aaDMs.
That is to say, the bits 0 and 1 that contribute to 
constructing a cryptographic key are encoded into a physical
system---a photon, say---via $\rho_0$ and $\rho_1$. Likewise,
the bit string $z = z_1z_2 \ldots z_n$ is represented by
$n$ different photons, the $i$th photon being in state $\rho_i$.
Thus the combined state for the string $z$ is given by
\BEqn
\rho_z = \rho_{z_1} \Tensor \rho_{z_2}\Tensor\cdots\Tensor\rho_{z_n}
\EEqn
where $\Tensor$ stands for the tensor product.

Now let $Z_0^{(n)}$ denote all the strings of length $n$ with
even parity (\ie with overall exclusive-or equal to \Zero)
and $Z_1^{(n)}$ all strings of length $n$ with odd parity.
Then define
\BEqn
\rho_j^{(n)} = 1/2^{n-1} \sum_{z \in Z_j^{(n)}} \rho_z
\EEqn
for $j = 0,1$.
In \BMS these two \aaDMs
and explicitly calculated in order to compute their
\aaSD as a function
of $n$ and $\alpha$.
This is extremely important because the parity bit appears in
the proof of security \MyCit{BihamBoBrGrMo97} of the BB84 key
exchange protocol\MyCit{BennettBr84}.

Here we compute the distinguishability between $\rho_0^{(n)}$ and
$\rho_1^{(n)}$ in terms of \aaKD and Bhattacharyya coefficient.
For the special case $n=2$ we also study the inequalities obtained in
Theorem \MyRef{Th:Ineqs}, as an illustration of how tight the
bounds are.  Observe that, at this point in time, the problem of the
parity bit is one of the few non-trivial (i.e., multi-dimensional
Hilbert-space) examples for which the \aaSD, \aaKD (and related
\aaPOE) and Bhattacharyya coeffecient can be computed.
For the next few paragraphs the reader is advised
to consult \BMS, or to take Eqs.~\RefEq{ParK}, \RefEq{ParB},
and \RefEq{ParMI} below as given.

First let us calculate the \aaKD $\QK( \rho_0^{(n)}, \rho_1^{(n)} )$
as a function of $n$ and $\alpha$.  In \BMS it is shown that 
$\Delta^{(n)} = \Half (\rho_0^{(n)} - \rho_1^{(n)})$ has non-zero
entries only on the secondary diagonal.  Moreover, it is not
difficult to see that all these entries equal $c^ns^n$, where
$c = \cos \alpha, s = \sin \alpha$. Therefore $\Delta^{(n)}$ has
$2^{n-1}$ eigenvalues equal to $- c^ns^n$, and $2^{n-1}$ eigenvalues
equal to $+ c^ns^n$, so
\BEqn \LabelEq{ParK}
\QK ( \rho_0^{(n)}, \rho_1^{(n)} ) = \sum \Abs{\lambda_j} =
|2 c s|^n = |\sin 2 \alpha|^n = |S|^n,
\EEqn
where $S = \sin 2 \alpha$.
Clearly, $ \LAcc \rho_0^{(n)} \RAcc$ and $ \LAcc \rho_1^{(n)} \RAcc $
are \aaEyI for all values of $\alpha\ne\pi/2$.  (Note that \BMS
proved exponential indistinguishability only for the case that
$\alpha\approx0$.)

Computing the Bhattacharrya coefficient between $\rho_0^{(n)}$ and
$\rho_1^{(n)}$ is a more elaborate calculation.  In \BMS Eqs.~(19)
and (20) it is shown that, with a minor change of basis,
$\rho_0^{(n)}$ and $\rho_1^{(n)}$ can be taken to be block-diagonal
with $2 \times 2$ blocks.  The $2\times2$ blocks are of the form
\BEqn
\sigma_0^{(n,k)}=\left(
\begin{array}{cc}
c^{2(n-k)}s^{2k} & c^ns^n \\[10pt]
c^ns^n & c^{2k}s^{2(n-k)}
\end{array}
\right)\mathrm{\ for\ even\ parity;\ }
\EEqn
and
\BEqn
\sigma_0^{(n,k)}=\left(
\begin{array}{cc}
c^{2(n-k)}s^{2k} & -c^ns^n \\[10pt]
-c^ns^n & c^{2k}s^{2(n-k)}
\end{array}
\right)\mathrm{\ for\ odd\ parity},
\EEqn
where $k$ ranges between $0$ and $n$.
For each $0\le k\le \lfloor n/2\rfloor$, the blocks
$\sigma_p^{(n,k)}$ and $\sigma_p^{(n,n-k)}$ each make an appearance
a total of $\Half\BinC{n}{k}$ times.

With this as a starting point, let us develop a convenient notation.
If $\sigma$ is an $n\times n$ positive semi-definite matrix of the
form
\BEqn
\left(
\begin{array}{cc}
\sigma^u & {\bf 0}_{pq}\\[10pt]
{\bf 0}_{qp} & \sigma^l
\end{array}
\right)\;,
\EEqn
where $\sigma^u$ is a $p\times p$ matrix, $\sigma^l$ is a $q\times q$
matrix, ${\bf 0}_{pq}$ is a $p\times q$ matrix, and $n=p+q$, then
we shall write this as $\sigma = \sigma^u \oplus \sigma^l$.
In this fashion, we have
\BEqn
\rho_0^{(n)}=\bigoplus_{k=1}^{2^{n-1}}\,\rho_{(0,k)}\;,
\EEqn
for the appropriate $2\times2$ matrices $\rho_{(0,k)}$.  Similarly
for $\rho_1^{(n)}$.

It is not difficult to see that the following three equalities hold:
\BEqnA
\Trace(\sigma^u \oplus \sigma^l ) &=& \Trace(\sigma^u) +
\Trace(\sigma^l) \\
(\sigma_0^u \oplus \sigma_0^l )(\sigma_1^u \oplus \sigma_1^l ) &=&
(\sigma_0^u  \sigma_1^u) \oplus (\sigma_0^u  \sigma_1^u) \\
\Sqrt{\sigma^u \oplus \sigma^l} &=& \Sqrt{\sigma^u} \oplus
\Sqrt{\sigma^l}
\EEqnA
{}From this it follows that
\BEqn
\Trace\sqrt{\sqrt{\sigma_0}\sigma_1\sqrt{\sigma_0}}\,=\,
\Trace\sqrt{\sqrt{\sigma_0^u}\sigma_1^u\sqrt{\sigma_0^u}}\,+\,
\Trace\sqrt{\sqrt{\sigma_0^l}\sigma_1^l\sqrt{\sigma_0^l}}\;,
\EEqn
which we can write in a short-hand notation as\footnote{Note that the
expressions in this short-hand version are not proper Bhattacharyya
coefficients: they are not normalized properly.}
\BEqn
\QB ( \sigma_0, \sigma_1 ) =
\QB ( \sigma_0^u, \sigma_1^u) + \QB ( \sigma_0^l, \sigma_1^l ).
\EEqn
Thus we can evaluate $\QB( \rho_0^{(n)}, \rho_1^{(n)} )$ by
evaluating each block individually and summing the results.
In particular, we find that
\BEqn
\QB\!\left( \sigma_0^{(n,k)}, \sigma_1^{(n,k)}\right) =
\QB\Big( \sigma_0^{(n,n-k)}, \sigma_1^{(n,n-k)}\Big) =
\Big|c^{2(n-k)} s^{2k} - c^{2k} s^{2(n-k)}\Big|.
\EEqn
Summing up over all blocks of $\rho_i^{(n)}$ we get
\BEqn
\QB ( \rho_0^{(n)}, \rho_1^{(n)} ) = 
\sum_{k=0}^{\lfloor n/2 \rfloor} \BinC{n}{k}
\left|c^{2(n-k)} s^{2k} - c^{2k} s^{2(n-k)}\right|.
\EEqn

For the case $n=2$ this expression reduces to
\BEqn \LabelEq{ParB}
\QB ( \rho_0^{(2)}, \rho_1^{(2)} ) = \left|c^4 - s^4\right| =
\left|(c^2 - s^2)(c^2 + s^2)\right| = |C| ,
\EEqn
where $C = \cos 2\alpha$.

For the \aaSD in the special case $n=2$, \BMS Eq.~(44) gives
\BEqn \LabelEq{ParMI}
\QSD  ( \rho_0^{(2)}, \rho_1^{(2)} ) =
\frac{1}{2} (1 + C^2)
 I_2\!\left(\frac{C^2}{1+C^2}\right) + \frac{S^2}{2} \ .
\EEqn

\begin{figure}[t]
\BCenter
\epsfig{file=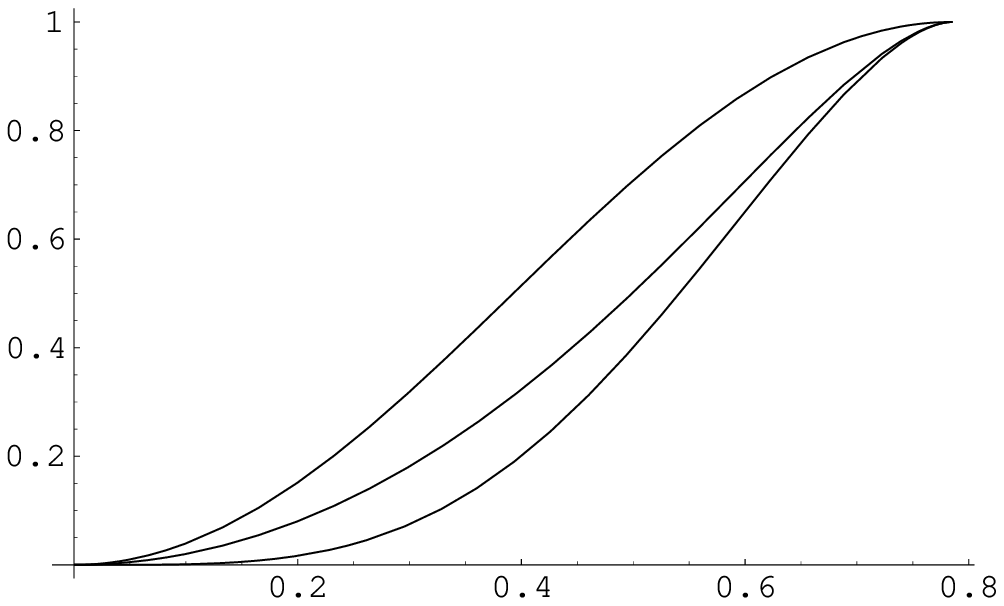,height=7cm}
\ECenter
\caption{\small
Equation \RefEq{QPELeqSD} for the parity bit with $n=2$
and with $\alpha \in [0,\pi/4]$ on the horizontal axis.
}
\MyLabel{fig:qind2}
\end{figure}

We are now in a position to substitute Eqs.~\RefEq{ParK},
\RefEq{ParB} and \RefEq{ParMI}
into Eqs.~\RefEq{QBLeqK}, \RefEq{QPELeqSD} and \RefEq{QBLeqSD}.
Observe that Eq.~\RefEq{QBLeqK} holds automatically,
in fact with equality on the right hand side.
Equations \RefEq{QPELeqSD} and \RefEq{QBLeqSD} are illustrated
in Figure \MyRef{fig:qind2} and \MyRef{fig:qind3} respectively.
The horizontal axis represents the angle $\alpha$ between
$\Ket{\psi_i}$ and $\Vec{1}{0}$,
meaning that for $\pi/4 \ (\approx 0.785)$ the states
$\Ket{\psi_0}$ and $\Ket{\psi_1}$ are orthogonal.
The fact that the bounds based on the Bhattacharyya coefficient
are fairly tight can be explained by the fact that the function
$2\Sqrt{x(1-x)}$ resembles $\BShE(x)$ quite well.

\subsection{Acknowledgments}

We are grateful to Tal Mor for helping us realize the usefulness of
these results.  Special thanks are also due Michel Boyer and Claude
Cr\'epeau.

CAF was supported by a Lee A. DuBridge Fellowship and by DARPA through
the Quantum Information and Computing (QUIC) Institute administered by
the US Army Research Office.
JvdG was supported by Canada's NSERC and Qu\'ebec's FCAR.

\begin{figure}[t]
\BCenter
\epsfig{file=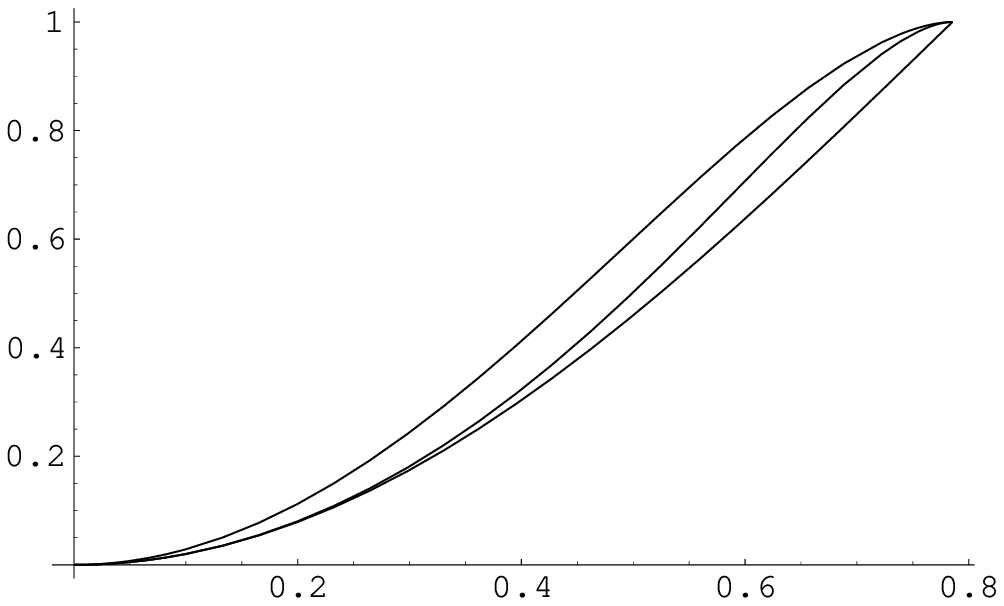,height=7cm}
\ECenter
\caption{\small
Equation \RefEq{QBLeqSD} for the parity bit with $n=2$
and with $\alpha \in [0,\pi/4]$ on the horizontal axis.
}
\MyLabel{fig:qind3}
\end{figure}


\begin{thebibliography}{10}

\bibitem{BihamBoBrGrMo97}
E.~Biham, M.~Boyer, G.~Brassard, J.~van~de Graaf, and T.~Mor, {\em Security of
  Quantum Key Distribution Against All Collective Attacks}.
\newblock 1997.
\newblock in preparation.

\bibitem{GoldwasserMiRa89}
S.~Goldwasser, S.~Micali, and C.~Rackoff, ``The knowledge complexity of
  interactive proof-systems,'' {\em SIAM. J. Computing}, vol.~18, pp.~186--208,
  Feb. 1989.

\bibitem{Beaver91}
D.~Beaver, ``Secure multiparty protocols and zero-knowledge proof systems
  tolerating a faulty minority,'' {\em Journal of Cryptology}, vol.~4, no.~2,
  pp.~75--122, 1991.

\bibitem{FuchsPhD}
C.~A. Fuchs, {\em Distinguishability and Accessible Information in Quantum
  Theory}.
\newblock PhD thesis, University of New Mexico, 1995.

\bibitem{BennettMoSm96}
C.~Bennett, T.~Mor, and J.~Smolin, ``The~parity bit in quantum cryptography,''
  {\em Physical Review~A}, vol.~54, no.~3, p.~2675, 1996.
\newblock also available at \WebPage{http://xxx.lanl.gov/ps/quant-ph/9604040}.

\bibitem{HughesBook}
R.~I.~G. Hughes, {\em The structure and interpretation of quantum mechanics}.
\newblock Harvard University Press, 1989.

\bibitem{IshamBook}
C.~Isham, {\em Lectures on Quantum Theory}.
\newblock Imperial College Press, 1995.

\bibitem{BrassardCrJoLa93}
G.~Brassard, C.~Cr\'epeau, R.~Jozsa, and D.~Langlois, ``A quantum bit
  commitment scheme provably unbreakable by both parties,'' in {\em Proc.\
  $34$th IEEE Symp.\ on Foundations of Comp.\ Science}, pp.~362--371, IEEE,
  1993.

\bibitem{SudberyBook}
Sudbery, {\em Quantum Mechanics and the Particles of Nature --- an outline for
  Mathematicians}.
\newblock Cambridge University Press, 1986.

\bibitem{PeresBook}
A.~Peres, {\em Quantum Theory: Concepts and Methods}.
\newblock Kluwer Academic Publishers, 1993.

\bibitem{Mermin96}
N.~D. Mermin, ``The {I}thaca {I}nterpretation of {Q}uantum {M}echanics,'' Tech.
  Rep. 9609013, LANL Quant-ph Archives, 1996.
\newblock Available at \WebPage{http://xxx.lanl.gov/ps/quant-ph/9609013}.

\bibitem{BenBassat82}
M.~Ben-bassat, ``Use of distance measures, information measures and error
  bounds in feature evaluation,'' in {\em Handbook of Statistics, Volume 2 ---
  Classification, Pattern Recognition and Reduction of Dimensionality} (P.~R.
  Krishnaiah and L.~N. Kanal, eds.), ch.~35, pp.~773--791, North-Holland, 1982.

\bibitem{HelstromBook}
C.~W. Helstrom, {\em Quantum Detection and Estimation Theory}.
\newblock Mathematics in Science and Engineering, vol.\ 123, Academic Press,
  1976.

\bibitem{Fuchs96}
C.~A. Fuchs, ``Information gain vs. state disturbance in quantum theory,'' {\em
  to appear in Fortschritte der Physik}, 1996.
\newblock available at \WebPage{http://xxx.lanl.gov/ps/quant-ph/9611010}.

\bibitem{Yao82b}
A.~Yao, ``Protocols for secure computations,'' in {\em Proc.\ $23$rd IEEE
  Symp.\ on Foundations of Comp.\ Science}, (Chicago), pp.~160--164, IEEE,
  1982.

\bibitem{GoldwasserMi84}
S.~Goldwasser and S.~Micali, ``Probabilistic encryption,'' {\em Journal of
  Computer and System Sciences}, vol.~28, pp.~270--299, Apr. 1984.

\bibitem{Toussaint72}
G.~T. Toussaint, ``Comments on `{T}he divergence and {B}hattacharyya distance
  measures in signal selection,'' {\em IEEE Transactions on Communication
  Technology}, vol.~COM-20, p.~485, 1972.

\bibitem{ReedSimonBook}
M.~Reed and B.~Simon, {\em Methods of modern mathematical physics --- part I:
  functional analysis}.
\newblock Academic Press, 1972.

\bibitem{Orlowski96}
A.~Or{\l}owski, ``Measures of distance between quantum states,'' in {\em
  Proceedings of the Fourth Workshop on Physics and Computation ---
  PhysComp~'96}, (Boston), pp.~239--242, New England Complex Systems Institute,
  1996.

\bibitem{Kailath67}
T.~Kailath, ``The divergence and {B}hattacharyya distance measures in signal
  selection,'' {\em IEEE Transactions on Communication Technology},
  vol.~COM-15, no.~1, pp.~52--60, 1967.

\bibitem{Toussaint71}
G.~T. Toussaint, ``Some functional lower bounds on the expected divergence for
  multihypothesis pattern recognition, communication, and radar systems,'' {\em
  IEEE Transactions on Systems, Man, and Cybernetics}, vol.~SMC-1,
  pp.~384--385, 1971.

\bibitem{FuchsCa95}
C.~A. Fuchs and C.~M. Caves, ``Mathematical techniques for quantum
  communication theory,'' {\em Open Systems and Information Dynamics}, vol.~3,
  no.~3, pp.~345--356, 1995.

\bibitem{Uhlmann76}
A.~Uhlmann, ``The `transition probability' in the state space of a
  $\,^*$-algebra,'' {\em Reports on Mathematical Physics}, vol.~9,
  pp.~273--279, 1976.

\bibitem{Jozsa94}
R.~Jozsa, ``Fidelity for mixed quantum states,'' {\em Journal of Modern
  Optics}, vol.~41, no.~12, pp.~2315--2323, 1994.

\bibitem{Shannon48}
C.~E. Shannon, ``A mathematical theory of communication,'' {\em Bell Sys.\
  Tech.\ J.}, vol.~27, pp.~623--656, 1948.

\bibitem{Lindley56}
D.~V. Lindley, ``On a measure of the information provided by an experiment,''
  {\em Ann. Math. Statist.}, vol.~27, pp.~986--1005, 1956.

\bibitem{FuchsCa94}
C.~A. Fuchs and C.~M. Caves, ``Ensemble-dependent bounds for accessible
  information in quantum mechanics,'' {\em Physical Review Letters}, vol.~73,
  no.~23, pp.~3047--3050, 1994.

\bibitem{MorPhD}
T.~Mor, {\em Quantum Memory in Quantum Cryptography}.
\newblock PhD thesis, Technion, Haifa, 1997.

\bibitem{HellmanRa70}
M.~E. Hellman and J.~Raviv, ``Probability of error, equivocation, and the
  {C}hernoff bound,'' {\em IEEE Transactions on Information Theory},
  vol.~IT-16, no.~4, pp.~368--372, 1970.

\bibitem{CoverTh91}
T.~M. Cover and J.~A. Thomas, {\em Elements of Information Theory}.
\newblock Wiley Series in Telecommunications, New York: John Wiley \& Sons,
  1991.

\bibitem{BennettBr84}
C.~H. Bennett and G.~Brassard, ``Quantum cryptography: Public key distribution
  and coin tossing,'' in {\em Proceedings of IEEE International Conference on
  Computers, Systems and \mbox{Signal} Processing}, (Bangalore, India),
  pp.~175--179, 1984.

\end{thebibliography}
\end{document}